\documentclass[12pt,twoside]{article}
\usepackage[mathscr]{eucal}
\usepackage{amsmath,amsfonts,amssymb,amsthm,mathabx,empheq}
\usepackage{times}
\usepackage{pdfsync}
\usepackage{cite}
\usepackage{url}
\usepackage{hyperref}
\usepackage{tensor}
\usepackage{color}
\usepackage{multicol}
\usepackage{bbold}

\advance\textheight\topskip
\allowdisplaybreaks[1]

\newcommand\td{\text{d}}
\newcommand\cO{{\cal O}}
\newcommand{\p}{\partial}

\newcommand{\be}{\begin{equation}}
\newcommand{\ee}{\end{equation}}
\newcommand{\bea}{\begin{eqnarray}}
\newcommand{\eea}{\end{eqnarray}}

\def\bz{\bar z}

\def\half{\frac12}

\def\bY{\bar Y}

\def\bp{\bar \partial}
\def\n{\nabla}
\def\bm{\bar{m}}
\def\bP{\bar P}
\def\bL{\bar{L}}
\def\bOmega{\bar\Omega}

\def\bA{\bar{A}}
\def\bB{\bar{B}}
\def\bL{\bar{L}}

\def \sd {\delta\hspace{-0.50em}\slash\hspace{-0.05em}}

\newcommand{\nn}{\nonumber}
\newcommand*\xbar[1]{%
  \hbox{%
    \vbox{%
      \hrule height 0.5pt 
      \kern0.3ex
      \hbox{%
        \kern-0.0em
        \ensuremath{#1}%
        \kern-0.0em
      }%
    }%
  }%
}

\allowdisplaybreaks[1]

\DeclareFontFamily{OT1}{rsfs}{} \DeclareFontShape{OT1}{rsfs}{m}{n}{
<-7> rsfs5 <7-10> rsfs7 <10-> rsfs10}{}
\DeclareMathAlphabet{\mycal}{OT1}{rsfs}{m}{n}

\hypersetup{
colorlinks=true,
linktoc=page,    
linkcolor=blue,
citecolor=blue,
urlcolor=blue,
colorlinks=true,
}

\begin{document}

\title{Twisting asymptotic symmetries and algebraically special vacuum solutions}

\author{Pujian Mao and Weicheng Zhao}
\date{}

\def\mytitle{Twisting asymptotic symmetries and algebraically special vacuum solutions}

\addtolength{\headsep}{4pt}

\begin{centering}

  \vspace{1cm}

  \textbf{\large{\mytitle}}

  \vspace{1cm}

  {\large Pujian Mao and Weicheng Zhao}

\vspace{0.5cm}

\begin{minipage}{.9\textwidth}\small \it  \begin{center}
     Center for Joint Quantum Studies and Department of Physics,\\
     School of Science, Tianjin University, 135 Yaguan Road, Tianjin 300350, China
 \end{center}
\end{minipage}

\end{centering}

\begin{center}
Emails:  pjmao@tju.edu.cn,\,zhaoweichengok@tju.edu.cn
\end{center}

\begin{center}
\begin{minipage}{.9\textwidth}
\textsc{Abstract}: In this paper, we study asymptotic symmetries and algebraically special exact solutions in the Newman-Penrose formalism. Removing the hypersurface orthogonal condition in the well studied Newman-Unti gauge, we obtain a generic asymptotic solution space which includes all possible origins of propagating degree of freedom. The asymptotic symmetry of the generalized system extends the Weyl-BMS symmetry by two independent local Lorentz transformations with non-trivial boundary charges, which reveals new boundary degrees of freedom. The generalized Newman-Unti gauge includes algebraically special condition in its most convenient form. Remarkably, the generic solutions satisfying the algebraically special condition truncate in the inverse power of radial expansions and the non-radial Newman-Penrose equations are explicitly solved at any order. Hence, we provide the most general algebraically special solution space and the derivation is self-contained in the Newman-Penrose formalism. The asymptotic symmetry with respect to the algebraically special condition is the standard Weyl-BMS symmetry and the symmetry parameters consist only the integration constant order. We present the Kerr solution and Taub-NUT solution in the generalized Newman-Unti gauge in a simple form.

\end{minipage}
\end{center}

\thispagestyle{empty}

\newpage
\tableofcontents

\section{Introduction}

Exact solutions are of fundamental importance in General relativity and other theories of gravity. Normally, the equations of motion of gravity theory are very complicated due to their highly non-linear property. Certain simplifications or approximations are essential in obtaining solutions. The Petrov classification \cite{Petrov:2000bs} provides a very powerful tool to classify exact solutions. Algebraically special condition will significantly reduce the system and exact solutions in the reduced system are much more tractable \cite{Stephani:2003tm}. While, if one cares more about the asymptotic behavior of the solution, Bondi and collaborators established an elegant framework to formulate the Einstein equation as a characteristic initial value problem \cite{Bondi:1962px,Sachs:1962wk}.  Meanwhile, Newman and Penrose (NP) construct a remarkable formalism to understand gravitational theory by means of four null tetrad bases \cite{Newman:1961qr}. Under the Newman-Unti (NU) gauge \cite{Newman:1962cia}, the results in Bondi-Sachs (BS) system \cite{Bondi:1962px,Sachs:1962wk} can be equivalently obtained in the NP formalism. The NU system is connected to the BS system by radial coordinate transformation \cite{Barnich:2011ty}. In principle, any asymptotically flat solution can be put into the BS or NU system in series expansion. However they are in some sense not always convenient for exact asymptotically flat solutions, nor some generalized systems \cite{Fletcher,Ciambelli:2020eba,Adami:2021nnf,Geiller:2022vto,Geiller:2024amx}. The reason of the inconvenience is from the choice of different null directions. The NU system is built on a hypersurface orthogonal null direction. While to characterize algebraically special solution, the principle null direction (PND) is the best choice for representing exact solutions. In this paper, we will present an asymptotic analysis in the NP formalism by removing the hypersurface orthogonal condition, which we refer to as generalized NU system. The main motivation is to include exact solutions, such as Kerr solution \cite{Kerr:1963ud} and Taub-NUT solution \cite{Taub:1950ez,Newman:1963yy} in a simple form.

The advantage of the NP formalism is that it manifests the hypersurface orthogonal condition and other geometric meanings of certain gauge choices. In other words, one can apply the gauge choice to reduce the system in a very direct way. The other advantage in particular for asymptotic analysis in the NP formalism is encoded in its first order nature. First order differential conditions in metric formalism become algebraic conditions in first order formalism and one needs, in principle, one less order of symmetry parameters and solutions for computing the charges. In the NP formalism, the gauge transformation consists of diffeomorphism and local Lorentz transformations. Removing the hypersurface orthogonal condition leads to independent Lorentz transformations, which extends the Weyl-BMS symmetry \cite{Barnich:2016lyg,Barnich:2019vzx,Freidel:2021fxf}. We refer to the new residual Lorentz symmetry as twisting asymptotic symmetry. The full asymptotic symmetry is complete in the sense that all the radial dependence of the symmetry parameters are fixed. There are non-trivial finite charges associated to the twisting asymptotic symmetries which indicates that they present new boundary degrees of freedom.

The obtained solution space includes all possible origins of the propagating degree of freedom, which consists of three parts encoded in the evolution of the asymptotic shear $\sigma^0$, the tetrad field $M^0$ which determines the twist, and the conformal factor of the two surface $P$.\footnote{The degree of freedom from $P$ is relevantly independent. A common choice is first to fix it as a round sphere or punctured complex plane \cite{Barnich:2016lyg,Barnich:2021dta}.} The reduction of our results to the NU system is transparent, which is simply given by turning off the tetrad field $M^0$. The transformation laws of those fields indicate that one can turn off two of them without losing any propagating degree of freedom. A more appreciated choice is to turn off the twist and keep the asymptotic shear \cite{Bondi:1962px,Sachs:1962wk,Newman:1962cia}. In the second part of this paper, we consider the alternative choice. The main motivation is precisely from exact solutions. In the other choice, one has the possibility to set the full shear tensor $\sigma$ to zero which incorporates with algebraically special condition. According to the Goldberg-Sachs theorem \cite{Goldberg}, for any algebraically special spacetime, one can turn off the NP variables $\kappa$ and $\sigma$, where the repeated PND is chosen as one of the real NP tetrad basis. This null direction is normally not hypersurface orthogonal for exact solutions such as Kerr or Taub-NUT solution. We implement the simplification from the algebraically special condition and perform a self-contained asymptotic analysis in the NP formalism. We find that the solution space has a remarkable simplification in the radial direction. In terms of series expansion in the complex expansion $\rho$ and $\xbar\rho$, the solution space is given in a truncated form. We exactly verify that the non-radial NP equations only provide constraints at their leading order which was proposed by Newman and Unti \cite{Newman:1962cia}. To the best of our knowledge, this has never been proven or verified elsewhere for a generic algebraically special solution space in the NP formalism. This truncation is new in the NP formalism, it is known in metric formalism, see e.g., in Chapter 27 and 29 of \cite{Stephani:2003tm}, and the original references therein. The motivation of our derivation in the NP formalism is that, apart from verifying the NU conjecture for algebraically special solution, some geometric conditions and relations are better appreciated from a self-contained NP analysis than translation from the metric formalism. Note also that the translation itself may not be straightforward. Our NP derivation is coordinate independent in the sense that we do not need any specific arrangement of the coordinates system. The only needed ingredient is the affine parameter of the null geodesic associated to the repeated PND. Our derivation has one technical difference than that in \cite{Stephani:2003tm}. We choose a different gauge condition by setting $\xbar\alpha^0+\beta^0=0$, while the choice of \cite{Stephani:2003tm} is $\tau^0=0$ in the NP convention. Though the choice $\tau^0=0$ may be simpler in part of the computation, our choice is for applying the extremely valuable $\eth$ operator. Due to the existence of twist, it is more convenient to generalize the $\eth$ operator by including a time derivative piece. We denote the generalized opertator as $\eth_n$. The commutator of the generalized operator $\eth_n$ includes also a twist piece. We perform the reduction of the asymptotic symmetry from the generic case to the algebraically special case. Remarkably, we recover the standard Weyl-BMS symmetry \cite{Barnich:2016lyg,Barnich:2019vzx,Freidel:2021fxf} and all the symmetry parameters consist of only the integration constant piece. This is another realization of the BMS symmetry, which is much more tractable considering the significant simplifications in the radial direction. We also present the Kerr solution and Taub-NUT solution in the generalized NU system. They are given in  a very simple form.

The organization of this paper is as follows. In the next section, we review the NP formalism and specify the conditions with a twist. In section \ref{solution}, we derive the asymptotic solution space with a twist. In section \ref{asg}, we compute the asymptotic symmetry of the generalized NU system and the associated twisting charges. In section \ref{ASS}, we apply the algebraically special condition in the asymptotic analysis. The solution space and asymptotic symmetries in closed form are obtained. Section \ref{Exactsolutions} presents the Kerr solution and Taub-NUT solution in the generalized NU system. We close with some comments on future directions in the last section. There is one Appendix which provides a parallel derivation of asymptotic symmetries in terms of finite transformations.


\section{NP formalism with twist}
\label{NP}

The NP formalism \cite{Newman:1961qr} is a tetrad formalism with two real null basis, denoted as $e_1=l=e^2,\;e_2=n=e^1$, and two complex null basis $\;e_3=m=-e^4,\;e_4=\bar{m}=-e^3$. The two complex basis are conjugate of each other to guarantee that the metric constructed from the tetrad is real. The null basis vectors are set to have the following orthogonality and normalization conditions,
\be\label{tetradcondition}
l\cdot m=l\cdot\bm=n\cdot m=n\cdot\bm=0,\quad l\cdot n=1,\quad m\cdot\bm=-1.
\ee
The spin connection is labelled by 12 complex scalars denoted by Greek symbols,
\be\label{coefficient}
\begin{split}
&\kappa=\Gamma_{311}=l^\nu m^\mu\nabla_\nu l_\mu,\;\;\pi=-\Gamma_{421}=-l^\nu \bar{m}^\mu\nabla_\nu n_\mu,\\
&\epsilon=\half(\Gamma_{211}-\Gamma_{431})=\half(l^\nu n^\mu\nabla_\nu l_\mu - l^\nu \bar{m}^\mu\nabla_\nu m_\mu),\\
&\tau=\Gamma_{312}=n^\nu m^\mu\nabla_\nu l_\mu,\;\;\nu=-\Gamma_{422}=-n^\nu \bar{m}^\mu\nabla_\nu n_\mu,\\
&\gamma=\half(\Gamma_{212}-\Gamma_{432})=\half(n^\nu n^\mu\nabla_\nu l_\mu - n^\nu \bar{m}^\mu\nabla_\nu m_\mu),\\
&\sigma=\Gamma_{313}=m^\nu m^\mu\nabla_\nu l_\mu,\;\;\mu=-\Gamma_{423}=-m^\nu \bar{m}^\mu\nabla_\nu n_\mu,\\
&\beta=\half(\Gamma_{213}-\Gamma_{433})=\half(m^\nu n^\mu\nabla_\nu l_\mu - m^\nu \bar{m}^\mu\nabla_\nu m_\mu),\\
&\rho=\Gamma_{314}=\bar{m}^\nu m^\mu\nabla_\nu l_\mu,\;\;\lambda=-\Gamma_{424}=-\bar{m}^\nu \bar{m}^\mu\nabla_\nu n_\mu,\\
&\alpha=\half(\Gamma_{214}-\Gamma_{434})=\half(\bar{m}^\nu n^\mu\nabla_\nu l_\mu - \bar{m}^\nu \bar{m}^\mu\nabla_\nu m_\mu).
\end{split}
\ee
 The Weyl tensor is represented by five complex scalars, denoted as $\Psi_0$, $\Psi_1$, $\Psi_2$, $\Psi_3$, $\Psi_4$,
\begin{align}
\Psi_0=-C_{1313},\;\;\Psi_1=-C_{1213},\;\;\Psi_2=-C_{1342},\;\;\Psi_3=-C_{1242},\;\;\Psi_4=-C_{2324}.\nn
\end{align}
We will deal with asymptotically flat vacuum solution in this work. So the Ricci tensor vanishes. We would refer to \cite{Chandrasekhar} for the other notations.

One can obtain the following relations from the orthogonality conditions and normalization conditions in \eqref{tetradcondition},
\be
\begin{split}\label{1}
&l^\nu \n_\nu l_\mu=(\epsilon+\bar \epsilon)l_\mu -\kappa \bm_\mu - \bar \kappa m_\mu,\\
&l^\nu \n_\nu n_\mu=-(\epsilon+\bar \epsilon)n_\mu + \bar\pi \bm_\mu + \pi m_\mu,\\
&l^\nu \n_\nu m_\mu=(\epsilon - \bar\epsilon) m_\mu - \kappa n_\mu + \bar\pi l_\mu,\\
&n^\nu \n_\nu l_\mu=(\gamma+\bar \gamma)l_\mu -\tau \bm_\mu - \bar \tau m_\mu,\\
&m^\nu \n_\nu l_\mu=(\beta+\bar \alpha)l_\mu -\sigma \bm_\mu - \bar \rho m_\mu.
\end{split}
\ee
In the asymptotic analysis in the NP formalism near null infinity, the NU gauge \cite{Newman:1962cia} is well appreciated, which is based on two assumptions that $l$ is tangent to null geodesics and is a hypersurface orthogonal null direction. Those two conditions have the following consequences to the NP variables, $\kappa=0$ and $\rho=\xbar\rho$. Then, one can use third and first classes of tetrad rotations to set $\epsilon=0=\pi$ and $\tau=\xbar\alpha+\beta$. Correspondingly, $l$ is tangent to null geodesics with affine parameter and is the gradient of a scalar field. The rest three null basis are parallely transported along $l$. One normally choose this affine parameter as radial coordinate $r$ and choose the scalar field as time coordinate $u$, hence $l=\frac{\p}{\p r}$ and $l=du$.

When we consider algebraically special solution, the repeated PNDs are usually not hypersurface orthogonal. Hence, exact solutions, such as Kerr solution, are not in a simple form when transformed into the NU gauge. If one adopts the repeated PND to construct the null basis by choosing it as $l$, the Goldberg-Sachs theorem \cite{Goldberg} yields that $\sigma=0=\kappa$. Then, one can turn off $\epsilon$ by a third class of tetrad rotation, which will not touch the conditions $\sigma=0=\kappa$. Hence, the PND $l$ is tangent to a null geodesic with affine parameter. A first class of null rotation can turn off $\pi$ without changing $\sigma=\kappa=\epsilon=0$. 

In this work, we will implement an asymptotic study which can incorporate both hypersurface orthogonal null direction and PND. We only impose gauge conditions $\pi=\kappa=\epsilon=0$, hence $l=\frac{\p}{\p r}$. Then the tetrad system can be constructed as
\be\label{gaugetetrad}
\begin{split}
&n=W\frac{\p}{\p u} + U \frac{\p}{\p r} + X^A \frac{\p}{\p x^A},\\
&l=\frac{\p}{\p r},\;\;\;\;\;\;m= M\frac{\p}{\p u} + \omega\frac{\p}{\p r} + L^A \frac{\p}{\p x^A}.
\end{split}
\ee
Though we specify a coordinate system $(u,r,x^A)$, it is important to point out that we do not have any delicate arrangement for the choice of the non-radial coordinates. The fall-off conditions are chosen as follows
\be\label{boundaryconditions}
\begin{split}
&W=1+\cO(r^{-1}),\quad X^A=\cO(r^{-1}),\quad U=\cO(r),\quad M=\cO(r^{-1}),\\  &L^z=\cO(r^{-2}),\quad L^{\bz}=\cO(r^{-1}),\quad \omega=\cO(r^{-1}), \quad \rho=-\frac1r+\cO(r^{-2}),\\
& Re(\rho)=-\frac{1}{r}+\cO(r^{-3}),\quad \sigma=\cO(r^{-2}), \quad \alpha=\cO(r^{-1}), \quad \beta=\cO(r^{-1})\\
&\xbar\alpha+\beta=\cO(r^{-2}),\quad \lambda=\cO(r^{-1}),\quad \mu=\cO(r^{-1}), \quad \tau=\cO(r^{-1}).
\end{split}
\ee
Those conditions can be reached for solution space by three classes of tetrad rotations and the combined coordinates transformation which preserves the gauge conditions $\kappa=\epsilon=\pi=0$,\footnote{$\sigma=0$ will be also preserved for algebraically special case. Hence, any algebraically special solution can be set into those fall-off forms as well.} see, e.g., the arrangement for some of fall-off conditions in \cite{Newman:1962cia}.
As directional derivatives, the basis vectors are designated by special symbols,
\begin{align}
D=l^\mu\p_\mu,\;\;\;\;\Delta=n^\mu\p_\mu,\;\;\;\;\delta=m^\mu\p_\mu.
\end{align}
The full vacuum NP equations with respect to conditions $\kappa=\epsilon=\pi=0$ are arranged as

\textbf{Radial equations}
\bea
&&D\rho =\rho^2+\sigma\xbar\sigma,\label{gR1}\\
&&D\sigma=(\rho +\xbar\rho)\sigma + \Psi_{0},\label{gR2}\\
&&D\alpha=\rho  \alpha + \beta \xbar \sigma  ,\label{gR3}\\
&&D\beta  =\alpha \sigma + \xbar\rho  \beta + \Psi_{1},\label{gR4}\\
&&D\tau =\tau \rho +  \xbar \tau \sigma   + \Psi_1 ,\label{gR5}\\
&&D\lambda=\rho\lambda + \xbar\sigma\mu ,\label{gR6}\\
&&D\mu =\xbar\rho \mu + \sigma\lambda + \Psi_{2},\label{gR7}\\
&&D\gamma=\tau \alpha +  \xbar \tau \beta  + \Psi_2,\label{gR8}\\
&&D\nu =\xbar\tau \mu + \tau  \lambda + \Psi_3,\label{gR9}\\
&&D\Delta= - (\gamma + \xbar\gamma) D + \xbar\tau \delta + \tau \xbar\delta,\\
&&D\delta= - (\xbar\alpha +\beta) D + \xbar\rho \delta + \sigma \xbar\delta,\\
&&D\Psi_1 - \xbar\delta \Psi_0 =  4 \rho \Psi_1 - 4\alpha \Psi_0,\label{gR15}\\
&&D\Psi_2 - \xbar\delta \Psi_1 =   3\rho \Psi_2  - 2 \alpha \Psi_1- \lambda \Psi_0,\label{gR16}\\
&&D\Psi_3 - \xbar\delta \Psi_2 =  2\rho \Psi_3 - 2\lambda \Psi_1,\label{gR17}\\
&&D\Psi_4 - \xbar\delta \Psi_3 = \rho  \Psi_4 + 2 \alpha \Psi_3 - 3 \lambda \Psi_2.\label{gR18}
\eea

\textbf{Non-radial  equations}
\bea
&&\Delta\lambda  = \xbar\delta\nu- (\mu + \xbar\mu)\lambda - (3\gamma - \xbar\gamma)\lambda + \nu ( 3\alpha + \xbar\beta -\xbar\tau ) - \Psi_4,\label{gH1}\\
&&\Delta\rho= \xbar\delta\tau- \rho\xbar\mu - \sigma\lambda  - \tau (\xbar\tau + \alpha -\xbar\beta) + (\gamma + \xbar\gamma)\rho  - \Psi_2 ,\label{gH2}\\
&&\Delta\alpha = \xbar\delta\gamma +\rho \nu - (\tau + \beta)\lambda + (\xbar\gamma -\xbar \mu)\alpha  + \gamma(\xbar\beta - \xbar \tau)  -\Psi_3 ,\label{gH3}\\
&&\Delta \mu=\delta\nu-\mu^2 - \lambda\xbar\lambda - (\gamma + \xbar\gamma)\mu   +  \nu (\xbar\alpha + 3\beta - \tau),\label{gH4}\\
&&\Delta \beta=\delta\gamma - \mu\tau + \sigma\nu + \beta(\gamma - \xbar\gamma -\mu) - \alpha\xbar\lambda + \gamma (\xbar\alpha + \beta -\tau) ,\label{gH5}\\
&&\Delta \sigma=\delta\tau - \sigma\mu - \rho\xbar\lambda - \tau (\tau -\xbar\alpha + \beta) + (3\gamma - \xbar\gamma)\sigma  ,\label{gH6}\\
&&\Delta \delta=\delta \Delta + \xbar\nu D + (\xbar\alpha + \beta -\tau) \Delta + (\gamma-\xbar\gamma -\mu)\delta - \xbar\lambda \xbar\delta,\label{gH7}\\
&&\delta\rho - \xbar\delta\sigma=\rho(\xbar\alpha + \beta) - \sigma (3\alpha - \xbar\beta)  + \tau (\rho - \xbar\rho) - \Psi_1 ,\label{gH9}\\
&&\delta\alpha - \xbar\delta\beta=\mu\rho - \lambda\sigma + \alpha\xbar\alpha + \beta\xbar\beta - 2 \alpha\beta + \gamma(\rho - \xbar\rho) - \Psi_2 ,\label{gH10}\\
&&\delta\lambda - \xbar\delta\mu= \nu(\rho - \xbar\rho)-\mu (\alpha + \xbar\beta) + \lambda (\xbar\alpha - 3\beta) - \Psi_3 ,\label{gH11}\\
&&\xbar\delta \delta - \delta \xbar\delta=(\xbar\mu-\mu) D + (\xbar\rho-\rho)\Delta + (\alpha -\xbar\beta) \delta + (\beta - \xbar\alpha)\xbar\delta,\label{gH12}\\ 
&&\Delta\Psi_0 - \delta \Psi_1 = (4\gamma -\mu)\Psi_0 - (4\tau + 2\beta)\Psi_1 + 3\sigma \Psi_2,\label{gH14}\\
&&\Delta\Psi_1 - \delta \Psi_2 = \nu\Psi_0 + (2\gamma - 2\mu)\Psi_1 - 3\tau \Psi_2 + 2\sigma \Psi_3 ,\label{gH15}\\
&&\Delta\Psi_2 - \delta \Psi_3 = 2\nu \Psi_1 - 3\mu \Psi_2 + (2\beta - 2\tau) \Psi_3 + \sigma \Psi_4,\label{gH16}\\
&&\Delta\Psi_3 - \delta \Psi_4 = 3\nu \Psi_2 - (2\gamma + 4\mu) \Psi_3 + (4\beta - \tau) \Psi_4.\label{gH17}
\eea


\section{Twisting solution space in NP formalism}
\label{solution}

The NP variables satisfying the NP equations in asymptotic expansions are given by
\begin{align}
&\Psi_0=\frac{\Psi_0^0(u,z,\bz)}{r^5}+\cO(r^{-6}),\quad \Psi_1=\frac{\Psi_1^0(u,z,\bz)}{r^4}-\frac{\xbar\eth_n \Psi_0^0 + i\Sigma \Psi_1^0}{r^5}+\cO(r^{-6}),\nn\\
&\Psi_2=\frac{\Psi_2^0(u,z,\bz)}{r^3}-\frac{\xbar\eth_n \Psi_1^0 + i\Sigma \Psi_2^0}{r^4}+\cO(r^{-5}),\quad \Psi_3=\frac{\Psi_3^0(u,z,\bz)}{r^2}-\frac{\xbar\eth_n\Psi_2^0 + i\Sigma \Psi_3^0}{r^3}+O(r^{-4}),\nn\\
&\Psi_4=\frac{\Psi_4^0(u,z,\bz)}{r}-\frac{\xbar\eth_n \Psi_3^0 + i\Sigma \Psi_4^0}{r^2}+\cO(r^{-3}),\nn\\
&\nn\\
&\rho=-\frac{1}{r}+\frac{i\Sigma(u,z,\bz)}{r^2}+\frac{\Sigma^2-\sigma^0\xbar\sigma^0}{r^3}+\cO(r^{-4}),\nn\\
&\sigma=\frac{\sigma^0(u,z,\bz)}{r^2}-\frac{(\Sigma^2 -\sigma^0\xbar\sigma^0)\sigma^0 + \frac12 \Psi_0^0}{r^4} + \cO(r^{-5}),\nn\\
&\alpha=\frac{\alpha^0(u,z,\bz)}{r}+\frac{\xbar\alpha^0\xbar\sigma^0-i\Sigma \alpha^0}{r^2}+\frac{\alpha^0\sigma^0\xbar\sigma^0 - \Sigma^2\alpha^0}{r^3}+\cO(r^{-4}),\nn\\
&\beta=-\frac{\xbar\alpha^0(u,z,\bz)}{r}-\frac{\alpha^0\sigma^0+i\Sigma\xbar\alpha^0}{r^2}-\frac{\xbar\alpha^0\xbar\sigma^0\sigma^0-\Sigma^2\xbar\alpha^0+\frac12\Psi_1^0}{r^3}+\cO(r^{-4}),\nn\\
& \tau=\frac{\tau^0(u,z,\bz)}{r}-\frac{\xbar\tau^0\sigma^0+i\Sigma\tau^0}{r^2}+\cO(r^{-3}),\nn\\
&\lambda=\frac{\lambda^0(u,z,\bz)}{r}- \frac{\mu^0\xbar\sigma^0  + i\Sigma\lambda^0}{r^2} + \cO(r^{-3}),\nn\\
&\gamma=\gamma^0(u,z,\bz) +\frac{\xbar\tau^0\xbar\alpha^0 - \tau^0 \alpha^0}{r}+ \frac{i\Sigma(\xbar\alpha^0 \xbar\tau^0 - \alpha^0 \tau^0 ) - \frac12 \Psi_2^0}{r^2} + \cO(r^{-3}),\\
&\mu=\frac{\mu^0(u,z,\bz)}{r}-  \frac{\sigma^0\lambda^0+\Psi^0_2-i\Sigma \mu^0}{r^2} + \cO(r^{-3}),\nn \\
&\nu=\nu^0(u,z,\bz) -\frac{\xbar\tau^0\mu^0 + \tau^0 \lambda^0 + \Psi^0_3}{r}+ O(r^{-2}),\nn\\
&\nn\\
&L^z=-\frac{\bP(u,z,\bz)\sigma^0}{r^2} + \cO(r^{-3}),\;\;\;\; L^{\bz}=\frac{P(u,z,\bz)}{r}+\frac{i\Sigma P }{r^2}+O(r^{-3}),\nn\\
&\bL^{\bz}=-\frac{P \xbar\sigma^0}{r^2} + \cO(r^{-3}),\quad \bL^{z}=\frac{\bP}{r}-\frac{i \Sigma \bP }{r^2}+\cO(r^{-3}),\nn\\
&M=\frac{M^0(u,z,\bz)}{r}-\frac{\sigma^0\xbar M^0 - i \Sigma M^0}{r^2} +  \cO(r^{-3}),\nn\\
&\omega=\frac{\omega^0(u,z,\bz)}{r}-\frac{\sigma^0\xbar\omega^0 + \frac12 \Psi_1^0 - i \Sigma \omega^0}{r^2}+\cO(r^{-3}),\nn\\
&X^z=-\frac{\tau^0\bP}{r}+O(r^{-2}),\quad W=1-\frac{\tau^0\xbar M^0 + \xbar\tau^0 M^0}{r}+\cO(r^{-2}), \nn\\
&U=-(\gamma^0+\xbar\gamma^0)r + U^0(u,z,\bz)\nn\\
&\hspace{2cm}- \frac{\frac12(\Psi_2^0 + \xbar\Psi_2^0) + (\tau^0\xbar\omega^0 + \xbar\tau^0\omega^0)  - 2 i\Sigma (\xbar\alpha^0 \xbar\tau^0 - \alpha^0 \tau^0) }{r} + \cO(r^{-2}),\nn
\end{align}
where
\begin{align}
&\alpha^0=\frac12 (\xbar M^0 \p_u \ln P + \bP \p \ln P),\quad \gamma^0=-\frac12 \p_u \ln \bP, \quad \tau^0=-\p_u M^0 - 2 M^0 \xbar\gamma^0,\nn\\
&\mu^0=2i\Sigma \gamma^0+\mu^0_r,\quad \mu_r^0=-\eth_n\alpha^0-\xbar\eth_n \xbar\alpha^0,\quad 2i\Sigma=\eth_n \xbar M^0 - \xbar \eth_n M^0, \nn\\
&\omega^0=2i\Sigma \tau^0- i\eth_n \Sigma +\xbar\eth_n \sigma^0, \quad \lambda^0=\p_u \xbar\sigma^0 + (3\gamma^0 - \xbar\gamma^0) \xbar\sigma^0+ (\xbar\tau^0)^2-\xbar\eth_n \xbar\tau^0 \nn\\
&\nu^0=\xbar\eth_n (\gamma^0+\xbar\gamma^0)-\xbar\tau^0(\gamma^0+\xbar\gamma^0), \quad \Psi_3^0=2i\Sigma \nu^0 + \xbar\eth_n \mu^0 - \eth_n \lambda^0,\nn\\
&\Psi_4^0=\xbar\eth_n \nu^0 - 4\gamma^0\lambda^0 - \xbar\tau^0\nu^0 - \p_u\lambda^0,\nn\\
&U^0=\xbar\eth_n\tau^0-i\Sigma(\gamma^0 + 3 \xbar\gamma^0) - i\p_u \Sigma + \mu_r^0 - \tau^0\xbar\tau^0,\nn\\
&\Psi_2^0-\xbar\Psi_2^0=i\Sigma (2U^0+\mu^0 + \xbar\mu^0) + \xbar\eth_n \omega^0 - \eth_n \xbar\omega^0+\xbar\sigma^0\xbar\lambda^0 - \sigma^0 \lambda^0,\nn
\end{align}
and
\begin{align}
&\p_u\Psi^0_0 + (\gamma^0 + 5 \xbar \gamma^0)\Psi^0_0=\eth_n\Psi^0_1+3\sigma^0\Psi^0_2 - 4\tau^0\Psi_1^0,\nn\\
&\p_u\Psi^0_1 + 2 (\gamma^0 + 2 \xbar \gamma^0)\Psi^0_1=\eth_n\Psi^0_2+2\sigma^0\Psi^0_3-3\tau^0 \Psi_2^0,\nn\\
&\p_u\Psi^0_2 + 3 (\gamma^0 + \xbar \gamma^0)\Psi^0_2=\eth_n\Psi^0_3 + \sigma^0\Psi^0_4-2\tau^0\Psi_3^0,\nn
\end{align}
as well as the identities
\begin{align}
&\p_u \alpha^0=-\xbar\eth_n \xbar\gamma^0 + \xbar\gamma^0 \xbar\tau^0 - 2 \gamma^0\alpha^0,\nn\\
&\p_u \mu^0=\eth_n \nu^0 - 2 \mu^0 (\gamma^0+\xbar\gamma^0)-\tau^0\nu^0,\nn\\
&\p_u\Psi^0_3 + 2 (2 \gamma^0 + \xbar \gamma^0)\Psi^0_3=\eth_n\Psi^0_4-\tau^0\Psi_4^0,\nn
\end{align}
which are satisfied automatically when the precise expressions of the NP variables are inserted. Note that $U^0$ is proven to be real. We define an operator $\eth_n$ which generalizes the extremely useful $\eth$ operator at null infinity \cite{Newman:1966ub} for a field of spin $s$ as,
\be
\begin{split}
\eth_n \eta^{(s)}&=(P\bp + M^0\p_u + 2 s \xbar\alpha^0)\eta^{(s)}\\
&=P\bP^{-s}\bp (\bP^s \eta^{(s)}) + M^0 \bP^{-s}\p_u (\bP^s \eta^{(s)}),
\end{split}
\ee
and
\be
\begin{split}
\xbar\eth_n \eta^{(s)}&=(\bP\p + \xbar M^0\p_u - 2 s \alpha^0)\eta^{(s)}\\
&=\bP P^{s}\p (P^{-s} \eta^{(s)}) + \xbar M^0 P^{s}\p_u (P^{-s} \eta^{(s)}),
\end{split}
\ee
where $(s)$ denotes the spin weight of the field $\eta^{(s)}$. The generalized operators $\eth_n$ and $\overline{\eth}_n$ have the same effect for raising and lowering the spin weight as $\eth$ and $\overline{\eth}$. They are the special case of the very generic eth operator defined in \cite{Penrose:1985bww} with restriction to the null infinity. The commutator of the generalized operators is
\be\label{commutator}
[\xbar\eth_n,\eth_n]\eta^{(s)} =-(2s\mu_r^0+2i\Sigma \p_u)\eta^{(s)}.
\ee
The spin weights of relevant fields are listed in Table \ref{t1}. 
\begin{table}[ht]
\caption{Spin weights}\label{t1}
\begin{center}\begin{tabular}{|c|c|c|c|c|c|c|c|c|c|c|c|c|c|c|c|c|c|c|c|c}
\hline
& $\eth_n$ & $\tau_0$ & $\alpha^0$ & $\gamma^0$ & $\nu^0$ & $\mu^0$ & $\sigma^0$ & $\lambda^0$  & $\Psi^0_4$ &  $\Psi^0_3$ & $\Psi^0_2$ & $\Psi^0_1$ & $\Psi_0^0$   \\
\hline
s & $1$& $1$& $-1$ &$0$& $-1$& $0$& $2$& $-2$  &
  $-2 $&  $-1$ & $0$ & $1$ & $2$  \\
\hline
\end{tabular}\end{center}\end{table}

To summarize, the solution space of the system is fully characterized by three arbitrary functions $\sigma^0$ $P$, and $M^0$ of three variables $(u,z,\bz)$ and initial data $\Psi_0$, $\Psi_1^0$, the real part of $\Psi_2^0$ of two variables $(u_0,z,\bz)$. 


\section{Twisting asymptotic symmetries and charges}
\label{asg}

As a tetrad system, the gauge transformation of the NP formalism is a combination of a diffeomorphism and a local Lorentz transformation. Let ${\xi}^\mu$ and ${{\omega}^{a}}_b=-{\omega_b}^a$
denote parameters for the two types of infinitesimal transformations, the gauge transformation of the NP variables are \cite{Barnich:2019vzx}
\begin{equation}
\begin{split}
&\delta_{\xi,\omega}{e_a}^\mu ={\xi}^\nu\partial_\nu
    {e_a}^\mu-\p_\nu{\xi}^\mu{e_a}^\nu +{\omega_a}^b{e_b}^\mu,\\
&\delta_{\xi, \omega} \Gamma_{a b c} = {\xi}^\nu \partial_\nu \Gamma_{a b c} - e_c^\mu \p_\mu {\omega}_{a b} + {\omega_a}^{d}\Gamma_{dbc}+ {\omega_b}^{d}\Gamma_{adc} + {\omega_c}^{d}\Gamma_{abd},\\
&\delta_{\xi, \omega} C_{abcd} ={\xi}^\nu \partial_\nu C_{abcd}
+ {{\omega}_a}^f C_{fbcd} + {{\omega}_b}^f C_{afcd}
+ {{\omega}_c}^f C_{abfd} + {{\omega}_d}^f C_{abcf}. 
\end{split}
\end{equation}
The ten gauge conditions yield
\begin{itemize}
\item
  $0=\delta_{\xi,\omega}\; e_1^u=-\p_r \xi^u \Longrightarrow
  \xi^u=f(u,z,\bar z)$.
\item
  $0=\delta_{\xi,\omega}\; \Gamma_{311}=-\p_r \omega^{24}+\sigma \omega^{23}+\rho\omega^{24}
  \Longrightarrow \omega^{24}= \frac{\omega^{24}_0(u,z,\bar z)}{r} - \frac{\omega^{24}_0 i \Sigma + \omega^{23}_0 \sigma^0}{r^2} + \cO(r^{-3})$.
\item
  $0=\delta_{\xi,\omega}\; \Gamma_{411}=-\p_r \omega^{23}+\xbar\rho\omega^{23} +\xbar\sigma \omega^{24}
  \Longrightarrow \omega^{23}= \frac{\omega^{23}_0(u,z,\bar z)}{r}+ \frac{\omega^{23}_0 i \Sigma - \omega^{24}_0 \xbar\sigma^0}{r^2} + \cO(r^{-3})$.
\item
  $0=\delta_{\xi,\omega}\; \Gamma_{211}=-\p_r \omega^{12}+(\alpha+\xbar\beta) \omega^{24}+(\xbar\alpha+\beta)\omega^{23}
  \Longrightarrow \omega^{12}= \omega^{12}_0(u,z,\bar z) + \cO(r^{-3})$.
\item
  $0=\delta_{\xi,\omega}\; e_1^r=- \p_r \xi^r +
  \omega^{2a}e_a^r \Longrightarrow \xi^r=-\omega^{12}_0 r + Z(u,z,\bar z) - \frac{\omega^{23}_0 \omega + \omega^{24}_0 \xbar\omega^0}{r^2} + \cO(r^{-2})$.
\item
  $0=\delta_{\xi,\omega}\; e_1^A=- \p_r \xi^A +
  \omega^{2a}e_a^A \Longrightarrow \xi^z=Y^z(u,z,\bar z) - \frac{\omega^{24}_0 \bP}{r} + \cO(r^{-2})$ and $\xi^{\bz}=Y^{\bz}(u,z,\bar z) - \frac{\omega^{23}_0 P}{r} + \cO(r^{-2})$.
\item
  $0=\delta_{\xi,\omega}\;
  \Gamma_{321}=- \p_r \omega^{14} - \xbar\lambda \omega^{23} - \xbar\mu \omega^{24}
  \Longrightarrow \omega^{14}=\omega^{14}_0(u,z,\bar z) +\frac{\omega^{24}_0\xbar\mu^0 + \omega^{23}_0 \xbar\lambda^0}{r} + \cO(r^{-2})$.
\item $0=\delta_{\xi,\omega}\;
  \Gamma_{421}=- \p_r \omega^{13} - \mu \omega^{23} - \lambda \omega^{24}
  \Longrightarrow
\omega^{13}=\omega^{13}_0(u,z,\bar z) +\frac{\omega^{23}_0\mu^0 + \omega^{24}_0 \lambda^0}{r}  + \cO(r^{-2})$.
\item
  $0=\delta_{\xi,\omega}\; \Gamma_{431}=- \p_r \omega^{34}  + (\xbar\alpha -\beta) \omega^{23} - (\alpha - \xbar\beta) \omega^{24} \Longrightarrow
  \omega^{34}=\omega^{34}_0(u,z,\bar z) + \frac{2\alpha^0 \omega^{24}_0 - 2 \xbar\alpha^0 \omega^{23}_0}{r} + \cO(r^{-1})$.
\end{itemize}
Note that all the omitted subleading orders of those symmetry parameters are completely fix once a solution of the NP system is given. The fall-off conditions yield
\begin{itemize}
\item
  $\delta_{\xi,\omega}e_2^u=\cO(r^{-1})
  \Longrightarrow \omega^{12}_0=\p_u f$.
\item $\delta_{\xi,\omega}\; e_2^A=\cO(r^{-1}) \Longrightarrow \p_u Y^A=0$.
\item $\delta_{\xi,\omega}\; e_3^z=\cO(r^{-2})
  \Longrightarrow \bar{\partial} Y^z=0$, hence $ Y^z = Y(z)$.
\item $\delta_{\xi,\omega}\; e_4^{\bz}=\cO(r^{-2})
  \Longrightarrow \partial Y^{\bz}=0$, hence $ Y^{\bz} = \bar{Y} (\bz)$.
\item
  $\delta_{\xi,\omega} \; (\Gamma_{314}+\Gamma_{413}) = \mathcal{O}(r^{-3})
  \Longrightarrow Z =\half (\eth_n \omega^{23}_0 + \xbar\eth_n \omega^{24}_0 - \omega^{23}_0\tau^0 - \omega^{24}_0\xbar\tau^0)$.
\item
  $\delta_{\xi,\omega} \; \Gamma_{213} =\mathcal{O}(r^{-2})
  \Longrightarrow \omega^{14}_0 = (\gamma^0 + \bar{\gamma}^0) \omega^{24}_0 - \eth_n \omega^{12}_0$.
\item
  $\delta_{\xi,\omega} \; \Gamma_{214} =\mathcal{O}(r^{-2})
  \Longrightarrow \omega^{13}_0 = (\gamma^0 + \bar{\gamma}^0) \omega^{23}_0 - \xbar\eth_n \omega^{12}_0$.
\end{itemize}
The parameters of residual gauge transformations are characterized by six functions
\begin{equation}
\begin{split}
&f(u,z,\bz),\quad Y^z=Y(z),\quad
  Y^{\bz}= \bY(\bz),\\
&\omega^{23}_0(u,r,z), \quad \omega^{24}_0(u,r,z),\quad \omega^{34}_0(u,r,z).
\end{split}
\end{equation}
As was done in \cite{Barnich:2019vzx}, it is useful to reparametrize residual gauge symmetries by trading the real function
$\p_uf(u, z,\bz)$ and the imaginary
$\omega^{34}_0(u,  z,\bz)$ for a complex parameter $\Omega(u, z,\bz)$ according to
\begin{equation}
\begin{split}
\partial_u f &= \half[\bar\p\bar Y-\bar Y\bar \p\ln (P\bar P) + \p Y
-Y\p \ln (P\bar P)] + f
(\gamma^0+\xbar\gamma^0)+\half(\Omega+\bOmega), \\
\omega^{34}_0 &= \half[\bar\p \bar Y-\bar Y\bar \p\ln P +\bar Y \bar \p
\ln \bar P- \p Y + Y\p \ln \bar P -Y \p \ln P]\\ &\hspace{8cm}+
f(\xbar\gamma^0 - \gamma^0)+
\half(\Omega-\bOmega).\\
\end{split}
\end{equation}
This redefinition of parameters simplifies the transformation on $P$ and $\bP$ as
\begin{equation}
\delta P = \Omega P ,\quad \delta \bar{P} = \bar{\Omega}\bar{P},
\end{equation}
which we refer to as a complex Weyl rescaling. It then determines for $f$ in terms of an
integration function $T(z,\bz)$ which characterizes the supertranslation, 
\begin{equation}
f(u, z,\bz) = \frac{1}{\sqrt{P\bar{P}}} [T(z,
\bar{z}) + \frac{\tilde{u}}{2} (\partial Y + \bar{\partial}
\bar{Y}) - Y \partial \tilde{u} - \bar{Y} \bar{\partial} \tilde{u} +
\frac{1}{2} (\tilde{\Omega} +\tilde{ \bar{\Omega}})],
\end{equation}
where
\begin{equation}
\tilde{u} = \int^u_{u_0} du' \sqrt{P\bar{P}}, \quad \tilde{\Omega} =
\int^u_{u_0} du' \sqrt{P\bar{P}}\,\Omega.
\end{equation}
The new symmetries represented by $\omega^{23}_0(u,r,z)$ and $\omega^{24}_0(u,r,z)$ are explicitly from removing the hypersurface orthogonal condition. The asymptotic symmetry characterized by the parameters $(\xi,\omega)$
realize a symmetry algebra at ull infinity as
\begin{equation}
\begin{split}
& [\delta_{\xi_1,\omega_1},\delta_{\xi_2,\omega_2}]\phi^\alpha=
\delta_{\hat\xi,\hat\omega}\phi^\alpha,\\
& \hat\xi^\mu=[\xi_1,\xi_2]^\mu,\\
& {{(\hat{\omega})}_a}^b={\xi_1}^\rho\p_\rho{\omega_{2a}}^b+{\omega_{1a}}^c{\omega_{2c}}^b
-(1\leftrightarrow2),
\end{split}
\end{equation}
where $\phi^\alpha$ denotes an arbitrary field. In particular,
\begin{equation}
  \begin{split}
  &\hat f=Y_1^A\p_A f_{2}+ f_{1}\p_u f_2 -(1\leftrightarrow
  2),\\
  & \hat Y^A=Y_1^B\p_B Y^A_2-Y_2^B\p_B Y^A_1,\\
  &\hat \Upsilon=f_1 \p_u \Upsilon_2 + Y_1^A \p_A \Upsilon_2 - \p_u f_1 \Upsilon_2 - \Upsilon_1 \Lambda_2 - (1\leftrightarrow
  2),\\
   &\hat {\bar \Upsilon}=f_1 \p_u {\bar \Upsilon}_2 + Y_1^A \p_A {\bar \Upsilon}_2 - \p_u f_1 {\bar \Upsilon}_2 + {\bar \Upsilon}_1 \Lambda_2- (1\leftrightarrow
  2),\\
   &\hat\Lambda=f_1 \p_u \Lambda_2 + Y_1^A \p_A \Lambda_2 - (1\leftrightarrow
  2),
\end{split}
\end{equation}
where we define $\Upsilon=\omega^{23}_0$ and $\Lambda=\omega^{34}_0$ for notational brevity. The asymptotic symmetries have the following actions on the solution space
\begin{align}
&\delta M^0=f\p_u M^0 + Y^A \p_A M^0 + M^0 \p_u f + \Lambda M^0 + \xbar\Upsilon - \eth_n f,\nn\\
&\delta \Sigma=f \p_u \Sigma + Y^A \p_A \Sigma - \frac{i}{2}(\eth_n \xbar \Upsilon - \xbar\eth_n \Upsilon + \Upsilon \tau^0 - \xbar\Upsilon \xbar\tau^0),\nn\\
&\delta \tau^0=f \p_u \tau^0 + Y^A \p_A \tau^0 + \tau^0 \p_u f + \Lambda \tau^0 - \p_u \xbar\Upsilon - 2\xbar\gamma^0 \xbar\Upsilon + \eth_n \p_u f,\nn\\
&\delta \sigma^0=f\p_u \sigma^0 + Y^A \p_A \sigma^0 + \sigma^0 \p_u f + 2 \Lambda \sigma^0 + \xbar\Upsilon \tau^0 - \eth_n \xbar\Upsilon ,\nn \\
&\delta \Psi_0^0= f\p_u \Psi_0^0 + Y^A \p_A \Psi_0^0 + 3\Psi_0^0 \p_u f + 2 \Lambda \Psi_0^0 + 4 \xbar\Upsilon \Psi_1^0,\nn \\
&\delta \Psi_1^0=  f\p_u \Psi_1^0 + Y^A \p_A \Psi_1^0 + 3\Psi_1^0 \p_u f +  \Lambda \Psi_0^0 + 3 \xbar\Upsilon \Psi_2^0,\nn \\
&\delta \Psi_2^0 = f\p_u \Psi_2^0 + Y^A \p_A \Psi_2^0 + 3\Psi_2^0 \p_u f + 2 \xbar\Upsilon \Psi_3^0,\nn 
\end{align}
It is clear from the transformation law that one can turn off $M^0$ by the residual Lorentz transformation characterized by $\xbar\Upsilon$. Correspondingly, the twist vanishes. Indeed, this can always be done near the null infinity. Namely, one can always have a hypersurface orthogonal null direction to construct the tetrad system. This is because the null infinity has a very universal structure for any asymptotically flat spacetime. The whole framework is based on the existence of a null infinity and the solution can be considered as flowing from null infinity to the bulk in the form of series expansion. However, for certain interesting exact solutions, it is very hard to do so. The main difficulty is from the proof of the integrability condition for transforming into a hypersurface orthogonal null direction. While, series expansion in $\frac1r$ significantly simplifies the radial integration. 

The three arbitrary functions $\sigma^0$ $P$, and $M^0$ can all represent propagating degree of freedom. One can turn off two of them by residual gauge transformations which still keeps the full propagating degree of freedom. Regarding to the transformation laws, one can find that turning off both $\sigma^0$ and $M^0$ reduces the supertranslation to ordinary translations \cite{Barnich:2013axa}. Hence, the common choice at null infinity is to turn off $M^0$ and fix $P$ to be a round sphere \cite{Sachs:1962wk,Sachs:1962zza}.

Normally, the next step is to compute the charges associated to the asymptotic symmetries, which will characterize the boundary degrees of freedom. As was stated in the introduction, our main motivation of the present work for generalizing the NU system is to incorporate with exact solutions and clarify the interplay between hypersurface orthogonal null direction and PND. We will not demonstrate the full charge analysis here but only compute the charges associated to the symmetries associated to the independent Lorentz transformation, including the twisting asymptotic symmetries and the imaginary part of the Weyl rescaling since they are the additions to the well studied Weyl-BMS symmetry \cite{Barnich:2019vzx,Freidel:2021fxf}.\footnote{Note that the imaginary part of the Weyl rescaling was revealed in \cite{Barnich:2019vzx}, the charge associated to this symmetry has not been computed. While the charge of the real part of the Weyl rescaling was computed in \cite{Freidel:2021fxf} for a $u$-indepdent Weyl factor of the boundary metric.} This will confirm that removing the hypersurface orthogonal arises new boundary degrees of freedom.

It is very efficient to use the form language to compute boundary charges for the NP formalism \cite{Liu:2022uox,Mao:2022ldv}.
The exact formulas of the forms $e_a$ are given by the co-tetrad system. In asymptotic form, they are
\begin{align}
&l=du - \frac{\xbar M^0}{\bP} dz - \frac{M^0}{P} d\bz + \cO(r^{-2}),\nn\\
&n=\left[(\gamma^0+\xbar\gamma^0) r - U^0\right] du + dr - \left[\frac{\xbar M^0(\gamma^0+\xbar\gamma^0) r }{\bP} + \frac{\xbar\omega^0 - \xbar M^0 U^0}{\bP}\right] dz \nn\\
&\hspace{3cm} - \left[\frac{ M^0(\gamma^0+\xbar\gamma^0) r }{P} + \frac{\omega^0 -  M^0 U^0}{P}\right] d\bz  + \cO(r^{-1}),\label{co-tetrad}\\
&m=-\tau^0 du + \left(-\frac{r}{\bP} + \frac{ \xbar M^0 \tau^0 - i\Sigma}{\bP}\right) dz + \frac{M^0 \tau^0 - \sigma^0}{P} d\bz + \cO(r^{-1}),\nn\\
&\bm=-\xbar\tau^0 du + \frac{\xbar M^0 \xbar\tau^0 - \xbar\sigma^0}{\bP} d z  + \left(- \frac{r}{P} + \frac{ M^0 \xbar\tau^0 + i\Sigma }{ P}\right) d\bz + \cO(r^{-1}).\nn
\end{align}
The connection one forms $\Gamma_{ab}$ are given as
\bea
&& \Gamma_{12}=-(\gamma + \xbar\gamma) l + \xbar\tau m  +\tau \bm,\\
&&\Gamma_{13}= -\tau l + \rho m + \sigma \bm,\\
&&\Gamma_{23}=\xbar\nu l - \xbar\mu m - \xbar\lambda \bm ,\\
&&\Gamma_{34}=(\gamma -\xbar\gamma) l - (\alpha-\xbar\beta) m  + (\xbar\alpha- \beta ) \bm.
\eea
The boundary charge from the Palatini action is defined by \cite{Godazgar:2020gqd,Godazgar:2020kqd,Godazgar:2022foc,Liu:2022uox,Mao:2022ldv}
\be\label{palatinicharge}
\sd {\cal H}_{Pa}= \frac{1}{32\pi G} \epsilon^{abcd} \int_{\partial \Sigma} \left[\delta (i_\xi \Gamma_{ab}  e_c \wedge e_d) - i_\xi( \delta \Gamma_{ab}\wedge e_c\wedge e_d) - \delta(\omega_{ab} e_c\wedge e_d)\right],
\ee
for field dependent symmetry parameters where $\cal{S}$ can be any constant-$u$ two surface at the null infinity to evaluate the boundary charge. The imaginary part of the Weyl rescaling charge is integrable and is given by
\be\label{Lambda}
Q_\Lambda=\frac{1}{8\pi G} \int_{\cal{S}} \frac{\Lambda}{P \bP} (\xbar M^0 \omega^0 - M^0 \xbar\omega^0).
\ee
We have computed the full expression of the twisting charge. There is a divergent piece at order $r$,
\be
\sd Q_{\Upsilon,\bar\Upsilon}^{(r)}=\frac{r}{8\pi G} \int_{\cal{S}} Z \delta \frac{1}{P\bP},\quad \quad Z =\half (\eth_n \Upsilon + \xbar\eth_n \xbar\Upsilon - \Upsilon\tau^0 - \xbar\Upsilon\xbar\tau^0).
\ee
One can in general regularize the divergent boundary charges, see, e.g., relevant treatments in \cite{Compere:2018ylh,Ruzziconi:2020wrb,Freidel:2021fxf,Geiller:2021vpg,Geiller:2024amx}. Here, we will simply remove the divergence by choosing a reduction of the solution space, as the main motivation here is to manifest the degrees of freedom associated to the twist. We choose a fixed Weyl factor of the boundary metric. For simplicity, we set $P\bP=1$,  which yields the null infinity with topology $\mathbb{R}\times \mathbb{C}$. Hence,
\be
P=e^{i P_0(u,z,\bz)},\quad \quad \bP=e^{-i P_0(u,z,\bz)}.
\ee
This choice will turn off the real part of the Weyl rescaling and keep the imaginary part $\Lambda$. The twisting charge in the reduced case is obtained as
\begin{multline}\label{Upsilon}
\sd Q_{\Upsilon,\bar\Upsilon}^{(0)}=\frac{1}{8\pi G} \int_{\cal{S}} \delta \left[ (\Upsilon M^0 + \xbar\Upsilon \xbar M^0)  (M^0 \nu^0 + \xbar M^0\xbar \nu^0)\right]\\
+ Z \bigg[2 M^0 \delta \alpha^0 +2 \xbar M^0 \delta \xbar \alpha^0 
+ 2 (M^0 \xbar\gamma^0 \delta \xbar M^0 + \xbar M^0 \gamma^0 \delta M^0 )\\
+ 2 \bigg(\Sigma - i (\gamma^0 - \xbar \gamma^0 ) M^0 \xbar M^0 + i (M^0 \alpha^0 - \xbar M^0 \xbar \alpha^0) + \frac{i}{2} (\xbar M^0 \tau^0 - M^0 \xbar \tau^0)\bigg)\delta P_0 \bigg].
\end{multline}
The charge is not integrable in this form. One may consider to include a field-dependent redefinition of the symmetry parameters \cite{Adami:2020ugu,Alessio:2020ioh,Ruzziconi:2020wrb,Adami:2021sko,Adami:2021nnf,Adami:2022ktn} to make the charge integrable. But there are three dynamical fields in the charge $M^0$, $\xbar M^0$, and $P_0$. While there are only two symmetry parameters $\Upsilon$ and $\xbar\Upsilon$. The system is overdetermined. Note also that the $\Lambda$ charge involves $\sigma^0$. So there are more dynamical fields than symmetry parameters in this system. One can further reduce the solution space by choosing a non-dynamical $P_0$. Then it is possible to have the twist charge in an integrable form. Here, we sketch a naive way for achieving that. The charge can be written as
\be\label{integrable}
\sd Q_{\Upsilon,\bar\Upsilon}^{(0)}=\frac{1}{8\pi G} \int_{\cal{S}}  \hat \Upsilon \delta M^0 + \xbar{ \hat\Upsilon} \delta \xbar M^0,
\ee
where $\hat \Upsilon$, $\xbar{ \hat\Upsilon}$ can be derived from the algebraic equations by identifying \eqref{integrable} with \eqref{Upsilon}. However, it is not completely clear to us if one can inverse the relation to solve out $\Upsilon$, $\bar\Upsilon$ as there are derivatives of $\Upsilon$, $\bar\Upsilon$ in the parameter $Z$. Suppose that one can solve out $\Upsilon$, $\bar\Upsilon$ in terms of $\hat \Upsilon$, $\xbar{ \hat\Upsilon}$ and the dynamical fields $M^0$, $\xbar M^0$. Then, considering $\hat \Upsilon$, $\xbar{ \hat\Upsilon}$ as field independent parameters will make the twist charge integrable. An interesting feature of the twist and $\Lambda$ charges is that they consist of only news functions. The charges are not conserved and there is no flux-balance law constraining them. Once we reduce the solution space by choosing a non-dynamical and $u$-independent $P_0$, the twist charge vanishes and the $\Lambda$ symmetry is turned off. We find a similar type of charge in a three dimensional setup \cite{Alessio:2020ioh} where the charge is related to the Weyl anomaly coefficient \cite{Bilal:2008qx}.  

Clearly, the charge expressions \eqref{Lambda} and \eqref{Upsilon} indicate that the new boundary degrees of freedom are indeed associated to the relaxation of removing hypersurface orthogonal condition. Setting $M^0=0=\xbar M^0$ turns off both charges \eqref{Lambda} and \eqref{Upsilon}. Then one can turn off $P_0$ by trivial gauge transformation in the reduced solution space. A normal choice for a punctured complex plane null infinity is to set directly $P=\bP=1$ \cite{Barnich:2016lyg,Barnich:2021dta}. Here we show that the existence of twist arises interesting boundary degree of freedom associated to the imaginary part of the Weyl rescaling. Hence, the choice $P=\bP=1$ for a punctured complex plane null infinity with twist will lose physical degree of freedom.


\section{Algebraically special solutions}
\label{ASS}

In this section, we will manifest the algebraically special condition in the asymptotic analysis in the NP formalism. We will choose the repeated PND as the null basis $l$. Consequently, we will have extra gauge conditions $\sigma=\Psi_0=\Psi_1=0$.\footnote{Note that those conditions can not be reached by residual gauge transformation. They are extra conditions which can decrease propagating degrees of freedom. By default, those conditions exclude all algebraically general solutions.} We will show that there is a remarkable simplification in the solution space and asymptotic symmetry, giving the $r$-dependence of all NP variables and symmetry parameters in terms of simple rational functions of $r$. 

\subsection{Solution space}

A generic solution of $\rho$ for the radial NP equation is \cite{Adamo:2009vu}
\be
\rho=-\frac{1}{r+i\Sigma}.
\ee
Then the twist parameter $\Sigma$ will be propagated to everywhere of the solution space. But with this expression, the full solution space is given in a closed form. The solutions to the radial equations are given by
\begin{align}
    &\rho=-\frac{1}{r+i\Sigma},\quad \alpha=\frac{\alpha^0}{r+i\Sigma},\quad \beta=-\xbar\alpha,\quad \tau=\frac{\tau^0}{r+i\Sigma},\quad \lambda=\frac{\lambda^0}{r+i\Sigma},\nn\\
    &M=\frac{M^0}{r-i\Sigma},\quad \omega=\frac{\omega^0}{r-i\Sigma},\quad L^{\bz}=\frac{P}{r-i\Sigma},\nn\\
    &W=1-\frac{\xbar\tau^0 M^0}{r-i\Sigma}-\frac{\tau^0\xbar M^0}{r+i\Sigma},\quad X^z=-\frac{\tau^0\bP}{r+i\Sigma},\nn\\
    &\Psi_2=\frac{\Psi_2^0}{(r+i\Sigma)^3},\quad \mu=\frac{\mu^0}{r-i\Sigma}-\frac{r\Psi_2^0}{(r+i\Sigma)^2(r-i\Sigma)},\nn\\ 
    &\gamma=\gamma^0 + \frac{\xbar\tau^0\xbar\alpha^0}{r-i\Sigma} - \frac{\tau^0\alpha^0}{r+i\Sigma}-\frac{\Psi_2^0}{2(r+i\Sigma)^2},\nn\\ 
    &U=-(\gamma^0+\xbar\gamma^0)r+U^0-\frac{\Psi_2^0}{2(r+i\Sigma)}-\frac{\xbar\Psi_2^0}{2(r-i\Sigma)} - \frac{\tau^0\xbar\omega^0}{r+i\Sigma}-\frac{\xbar\tau^0\omega^0}{r-i\Sigma},\label{generaltypetwosolutionr}\\
    & \Psi_3=\frac{\Psi_3^0}{(r+i\Sigma)^2}+\frac{\Psi_3^1}{(r+i\Sigma)^3}+\frac{\Psi_3^2}{(r+i\Sigma)^4},\nn\\
    &\quad \quad \Psi_3^1=-\xbar\eth_n\Psi_2^0,\quad \quad \Psi_3^2=\frac32 \Psi_2^0 (i\xbar\eth_n \Sigma + \xbar\omega^0),\nn\\
    &\Psi_4=\frac{\Psi_4^0}{r+i\Sigma} +\frac{\Psi_4^1}{(r+i\Sigma)^2}+\frac{\Psi_4^2}{(r+i\Sigma)^3}+\frac{\Psi_4^3}{(r+i\Sigma)^4}+\frac{\Psi_4^4}{(r+i\Sigma)^5} ,\nn\\
    &\quad \quad \Psi_4^1=-\xbar\eth_n \Psi_3^0 ,\quad \quad \Psi_4^2=\frac32\lambda^0\Psi_2^0 + \Psi_3^0(i\xbar\eth_n \Sigma + \xbar\omega^0)-\frac12\xbar\eth_n\Psi_3^1,\nn\\
    &\quad \quad \Psi_4^3=\Psi_3^1(i\xbar\eth_n \Sigma + \xbar\omega^0)-\frac13 \xbar\eth_n \Psi_3^2,\quad\quad  \Psi_4^4=\Psi_3^2 (i\xbar\eth_n \Sigma + \xbar\omega^0),\nn\\
    &\nu=\nu^0-\frac{\Psi_3^0 + \tau^0\lambda^0}{r+i\Sigma}-\frac{\xbar\tau^0\mu^0}{r-i\Sigma}+\frac{\xbar\tau^0\Psi_2^0 }{2(r^2+\Sigma^2)}-\frac{\Psi_3^1}{2(r+i\Sigma)^2}-\frac{\Psi_3^2}{3(r+i\Sigma)^3}.\nn
\end{align}
Correspondingly, the co-tetrad system is given by
\begin{align}
&l=du - \frac{\xbar M^0}{\bP} dz - \frac{M^0}{P} d\bz,\nn\\
&n=\left[(\gamma^0+\xbar\gamma^0) r - U^0 + \frac{rK+J\Sigma}{r^2+\Sigma^2} \right] du + dr \nn\\
&\hspace{1.5cm} - \left[\frac{\xbar M^0}{\bP} \left((\gamma^0+\xbar\gamma^0) r - U^0 + \frac{rK+J\Sigma}{r^2+\Sigma^2}\right)+ \frac{\xbar\omega^0 }{\bP}\right] dz \nn\\
&\hspace{3cm} - \left[\frac{ M^0}{P} \left((\gamma^0+\xbar\gamma^0) r - U^0 + \frac{rK+J\Sigma}{r^2+\Sigma^2}\right)+ \frac{\omega^0 }{P}\right] d\bz,\\
&m=-\tau^0 du + \left(-\frac{r}{\bP} + \frac{ \xbar M^0 \tau^0 - i\Sigma}{\bP}\right) dz + \frac{M^0 \tau^0}{P} d\bz,\nn\\
&\bm=-\xbar\tau^0 du + \frac{\xbar M^0 \xbar\tau^0 }{\bP} d z + \left(-\frac{r}{P} + \frac{M^0 \xbar\tau^0 + i\Sigma }{ P}\right) d\bz ,\nn
\end{align}
where $K$ and $J$ are the real and imaginary parts of $\Psi_2^0$,
\be
K=\frac12(\Psi_2^0 + \xbar\Psi_2^0),\quad\quad J=-\frac{i}{2}(\Psi_2^0 - \xbar\Psi_2^0).\nn
\ee
All the NP variables are in closed form in the expansion of $\rho=-\frac{1}{r+i\Sigma}$ and $\xbar\rho=-\frac{1}{r-i\Sigma}$. Then, it is possible to solve all the non-radial equations to every order, which will be implemented explicitly. Starting from \eqref{gH9}, we obtain that
\be
\omega^0=2i\Sigma \tau^0- i\eth_n \Sigma.
\ee
From \eqref{gH6}, we obtain
\be\label{lambda0}
\lambda^0=(\xbar\tau^0)^2-\xbar\eth_n \xbar\tau^0.
\ee
A constraint of $\Psi_2^0$ is yielded by \eqref{gH15} as
\be\label{p}
\eth_n \Psi_2^0=3\tau^0\Psi_2^0.
\ee
From \eqref{gH10}, we get
\be
\mu^0=2i\Sigma \gamma^0+\mu^0_r,
\ee
where $\mu_r^0=-\eth_n\alpha^0-\xbar\eth_n \xbar\alpha^0$ is the real part. Eq. \eqref{gH11} yields
\be
\Psi_3^0=2i\Sigma \nu^0 + \xbar\eth_n \mu^0 - \eth_n \lambda^0.
\ee
From the $u$-component of \eqref{gH12}, one can derive that
\be
2i\Sigma=\eth_n \xbar M^0 - \xbar \eth_n M^0=-\frac{\left[P(\bP W - \xbar M^0 X^z)-M^0 \bP X^{\bz}\right]^2}{P\bP} l_{[u}\p_z l_{\bz]}.
\ee
It is clear that $l$ is hypersurface orthogonal once $\Sigma=0$. Note that $l_{[u}\p_r l_{z]}=0$ from the solution of the radial equations. From $\bz$-component of \eqref{gH12}, it is obtained that
\be
\alpha^0=\frac12 (\xbar M^0 \p_u \ln P + \bP \p \ln P).
\ee
The $r$-component of \eqref{gH12} fixed the imaginary part of $\Psi_2^0$ as
\be\label{c}
\Psi_2^0-\xbar\Psi_2^0=i\Sigma (2U^0+\mu^0 + \xbar\mu^0) + \xbar\eth_n \omega^0 - \eth_n \xbar\omega^0.
\ee
We continue with the $u$-component of \eqref{gH7}, which yields
\be
\tau^0=-\p_u M^0 - 2 M^0 \xbar\gamma^0. 
\ee
The $\bz$-component and $r$-component of \eqref{gH7} lead to
\be
\gamma^0=-\frac12 \p_u \ln \bP.
\ee
and
\be
\nu^0=\xbar\eth_n (\gamma^0+\xbar\gamma^0)-\xbar\tau^0(\gamma^0+\xbar\gamma^0).
\ee
From \eqref{gH2}, we obtain
\be
U^0=\xbar\eth_n\tau^0-i\Sigma(\gamma^0 + 3 \xbar\gamma^0) - i\p_u \Sigma + \mu_r^0 - \tau^0\xbar\tau^0.
\ee
One can prove that $U^0$ is real using the expression of $\tau^0$ and the commutator of generalized operator $\eth_n$ in \eqref{commutator}. From \eqref{gH16}, there is another constraint of $\Psi_2^0$,
\be\label{u}
\p_u \Psi_2^0 + 3 (\gamma^0 + \xbar\gamma^0) \Psi_2^0=\eth_n \Psi_3^0 - 2\tau^0 \Psi_3^0.
\ee
Eq. \eqref{gH1} yields
\be
\Psi_4^0=\xbar\eth_n \nu^0 - 4\gamma^0\lambda^0 - \xbar\tau^0\nu^0 - \p_u\lambda^0.
\ee
Finally, Equations \eqref{gH3}-\eqref{gH5} and \eqref{gH17} are identities and do not lead to new relation. In particular,
\be
\p_u \alpha^0=-\xbar\eth_n \xbar\gamma^0 + \xbar\gamma^0 \xbar\tau^0 - 2 \gamma^0\alpha^0,
\ee
and
\be
\p_u \mu^0=\eth_n \nu^0 - 2 \mu^0 (\gamma^0+\xbar\gamma^0)-\tau^0\nu^0.
\ee
The solutions of the non-radial equations recover the leading results of previous section with $\sigma^0=0$. Remarkably, we exactly prove that there is no new constraint from any subleading order. This has been suggested by Newman and Unti \cite{Newman:1962cia}. Here we give a direct verification for any algebraically special solution. To summarize, for any given $M^0$ and $P$, an exact algebraically special solution of vacuum Einstein equation is specified once a solution of the constraints for $\Psi_2^0$, namely Eqs. \eqref{p}, \eqref{c}, \eqref{u}, is obtained.

\subsection{Asymptotic symmetries and transformation law}

The extra conditions $\Psi_1=\Psi_0=\sigma=0$ yield that $\omega^{23}_0=0=\omega^{24}_0$. The asymptotic symmetries are in closed form as
\be
\begin{split}
&\xi^u=f(u,z,\bz),\quad Y^z=Y(z), \quad Y^{\bz}=\bY(\bz), \quad \xi^r=-\p_u f r, \\
&\omega^{12}=\p_u f,\quad \omega^{13}=-\xbar\eth_n \p_u f,\quad \omega^{14}=-\eth_n \p_u f,\quad \omega^{34}=\Lambda(u,z,\bz).
\end{split}
\ee
The symmetry algebra recovers precisely the Weyl-BMS algebra \cite{Barnich:2019vzx}. The asymptotic symmetries have the following transformation law on the solution space,
\begin{align}
&\delta M^0=f\p_u M^0 + Y^A \p_A M^0 + M^0 \p_u f + \Lambda M^0 - \eth_n f,\nn\\
&\delta \Sigma=f \p_u \Sigma + Y^A \p_A \Sigma ,\nn\\
&\delta \Psi_2^0 = f\p_u \Psi_2^0 + Y^A \p_A \Psi_2^0 + 3\Psi_2^0 \p_u f.\nn 
\end{align}
Note that the residual gauge transformation can not turn off $\Sigma$. Hence, the twist in this system must not be pure gauge from the point of view of both propagating and boundary degrees of freedom. It is amusing to notice that the inhomogeneous piece in the transformation of $M^0$ is only supertranslation relevant as $-\eth_n f$ which is in a similar form as the transformation law of $\sigma^0$ in the twist-free case \cite{Barnich:2013axa,Barnich:2019vzx}. The difference of them is only from their different spin weight. The shear of the generator of the null infinity $\lambda^0$ is completely fixed by $\tau^0$ in \eqref{lambda0}. Hence, in the algebraically special case, $\tau^0$ is effectively the news tensor of this system, which indicates the propagating degree of freedom. The NP variables $M^0$ and $\tau^0$ in the algebraically spacial case are the analogue of $\sigma^0$ and $\lambda^0$ of the twist-free case. 

The next step will be the computation of the asymptotic charges. As we have shown that the solution space and asymptotic symmetry parameters are all in truncated forms. We expect that the charges are in closed form which indicates that the BMS symmetry becomes symplectic symmetry similar to three dimensional theories \cite{Compere:2015knw}, see also \cite{Compere:2014cna}. Surprisingly, there are propagating degree of freedom in the present case. This will be the first example of enhancement from asymptotic symmetry or near horizon symmetry to symplectic symmetry in the system with propagating degree of freedom. We will present the full charge analysis elsewhere.

\section{Exact solutions with twist}
\label{Exactsolutions}

It is very important for the applications of asymptotic symmetry and the associated charges to have exact solutions in the proposed system. In this section, we will present the exact forms of Kerr and Taub-NUT solution in the generalized NU system. Both solutions are given in a very simple forms in the generalized NU gauge. 

\subsection{Kerr}

The Kerr line-element in the Kerr-Schild coordinate $(u,r,\theta,\phi)$ is
\begin{multline}
  \label{eq:115}
  ds^2=\frac{\vartriangle}{\varrho\bar\varrho}(d u-a\sin^2\theta d \phi)^2
  -\frac{\sin^2\theta}{\varrho\bar\varrho}[a d
  u - (r^2+a^2) d \phi]^2\\+
  2dr(d u-a\sin^2\theta d \phi)-\varrho\bar\varrho(d\theta)^2,
\end{multline}
where we define
\begin{equation}
\vartriangle=r^2-2Mr+a^2,\quad
 \varrho= r+ia\cos\theta,\quad \bar\varrho=r-ia\cos\theta.
\end{equation}
The inverse metric is given by
\begin{equation}
  \label{eq:97ab}
  g^{\mu\nu}=\frac{1}{\varrho\bar\varrho}\begin{pmatrix}
    -a^2\sin^2\theta & r^2+a^2 & 0 & -a \\
    r^2+a^2 & -\vartriangle & 0 & a \\
    0 & 0 & -1 & 0 \\
    -a & a & 0 & -1/\sin^2\theta
  \end{pmatrix}
\end{equation}
We choose the tetrad system as
\begin{equation}
\begin{split}
&l=\frac{\p}{\p r},\quad \quad n=\frac{\p}{\p u} + (-\frac12 + \frac{Mr}{\varrho \xbar \varrho}) \frac{\p}{\p r},\\
&m=\frac{e^{i\phi}}{\sqrt2 \bar\varrho}\left( i a \sin \theta \frac{\p}{\p u} - i a \sin \theta \frac{\p}{\p r} - \frac{\p}{\p \theta} + i \csc \theta \frac{\p}{\p \phi}\right). 
\end{split}
\end{equation}
The non-vanishing spin coefficients and Weyl scalars are
\begin{equation}
  \begin{split}
    & \rho = - \frac{1}{\varrho},\quad 
    \alpha=\frac{e^{-i\phi}\cot\frac{\theta}{2}}{2\sqrt2 \varrho},\quad \beta = - \xbar\alpha,\\
    & \mu=-\frac{1}{2\xbar\varrho} +\frac{Mr}{\bar\varrho\varrho^2},\quad
    \gamma=\frac{M}{2\varrho^2},\quad \nu=\frac{i a M e^{-i\phi} \sin \theta}{\sqrt2 \varrho^3},\\
 & \Psi_2=-\frac{M}{(\varrho)^3},\quad \Psi_3=-\frac{3 i a M e^{-i\phi} \sin \theta}{\sqrt2 \varrho^4},\quad \Psi_4=\frac{3a^2 M e^{-2i\phi} \sin^2 \theta}{\varrho^5}.
    \end{split}
\end{equation}
In the complex stereographic coordinates $(z,\bz)$, where $\theta=2\text{arccot}\sqrt{z\bz}$ and $\phi=-\frac{i}{2}\ln \frac{z}{\bz}$, the tetrad system becomes
\begin{equation}
\begin{split}
&l=\frac{\p}{\p r},\quad n=\frac{\p}{\p u} + (-\frac12 + \frac{Mr}{\varrho \xbar \varrho}) \frac{\p}{\p r},\\
&m=\frac{1}{ \bar\varrho}\left( \frac{ i a z}{P_S} \frac{\p}{\p u} - \frac{ i a z}{P_S} \frac{\p}{\p r}  + P_S \frac{\p}{\p \bz}\right), \quad\quad P_S=\frac{1+z\bz}{\sqrt2}. 
\end{split}
\end{equation}
The non-vanishing spin coefficients and Weyl scalars are now
\begin{equation}
  \begin{split}
    & \rho = - \frac{1}{\varrho},\quad 
    \alpha=\frac{\bz}{2\sqrt2 \varrho},\quad \beta = - \xbar\alpha,\\
    & \mu=-\frac{1}{2\xbar\varrho} +\frac{Mr}{\bar\varrho\varrho^2},\quad
    \gamma=\frac{M}{2\varrho^2},\quad \nu=\frac{i a M \bz}{P_S \varrho^3},\\
 & \Psi_2=-\frac{M}{\varrho^3},\quad \Psi_3=-\frac{3 i a M \bz}{P_S \varrho^4},\quad \Psi_4=\frac{6a^2 M \bz^2 }{P_S^2\varrho^5}.
    \end{split}
\end{equation}

\subsection{Taub-NUT}

The Taub-NUT solution in the complex stereographic coordinates adapted to the NP conventions is \cite{Griffiths:2009dfa}
\be\label{Taub-NUT-metric}
\td s^2 = f(r)\left(\td t + 2 i N \frac{z\td \bz - \bz \td z}{1+z\bz}\right)^2 - \frac{\td r^2}{f(r)} - (r^2+N^2)\frac{4\td z \td \bz}{(1+z\bz)^2},
\ee
where $f(r)=\frac{r^2 - 2 Mr - N^2}{r^2 + N^2}$. Here, $N$ is the NUT parameter and $M$ is the mass parameter. Defining
\be
\tilde u=t-r+2\sqrt{-M^2-N^2} \arctan \frac{r-M}{-M^2-N^2} + M \ln (r^2-2Mr-N^2),
\ee
as the new time coordinate, we choose the null bases for the metric \eqref{Taub-NUT-metric} as 
\be
\begin{split}
&l=\p_r,\quad \quad n= \p_{\tilde u} - \frac{f}{2} \p_r,\\
&m=\frac{i\sqrt2 N z}{r- iN}\p_{\tilde u} -  \frac{1+z \bz}{\sqrt2 (r- i N)}\p_{\bz},\\
&\bm=-\frac{i\sqrt2 N \bz}{r+ iN}\p_{\tilde u} - \frac{1+z \bz}{\sqrt2 (r + i N)}\p_z.
\end{split}
\ee
The non-zero NP variables are
\be
\begin{split}
&\rho=-\frac{1}{r+iN},\quad \alpha=-\frac{\bz}{2\sqrt2 (r+iN)},\quad \beta=-\xbar\alpha,\quad \gamma=\frac{M+iN}{2(r+iN)^2},\\  & \mu=- \frac{1}{2(r-iN)}+\frac{(M+iN)r}{(r+iN)^2(r-iN)},\quad 
\Psi_2=-\frac{M+iN}{(r + iN)^3}.
\end{split}
\ee
Both null bases $l$ and $n$ are repeated principal null directions without shear but with twist \cite{Griffiths:2009dfa}.


\section{Applications and future directions}

There are a number of interesting applications and open questions that should be addressed in the future. We list some of them here.
\begin{itemize}

\item The most direct one is to find new exact solutions of vacuum Einstein gravity from the three constraints of $\Psi_2^0$ for particular choice of $M^0$ and $P$.

\item It will be very meaningful to check if the investigation of the algebraically spacial solution can be extended to the case with a cosmological constant.

\item Another extension is to include electromagnetic fields or other types of matter fields.

\item It is interesting to see if one can proof some kind of uniqueness theorem from the truncated algebraically special solution space when further constraint is required, such as stationary or axial symmetric, see, e.g., \cite{Wu:2008yi} for relevant investigation from asymptotic point of view.

\item 
Another interesting direction is to find the most general asymptotic symmetry algebra at null infinity. The present work focuses on the relaxation with a twist. The resulting asymptotic Killing vector has a fixed $r$-component. Another interesting relaxation was proposed in \cite{Geiller:2022vto,Geiller:2024amx} by imposing the differential NU gauge condition where there is new symmetry arise from the  $r$-component of the asymptotic Killing vector. Constructing the metric from our tetrad assumption \eqref{gaugetetrad} indicates that the only gauge condition is $g_{rr}=0$. It seems that this gauge choice is more relaxed than \cite{Geiller:2022vto,Geiller:2024amx}. However the rest conditions are hidden in the tetrad and spin coefficient relations which represent the existence of special null directions that can not been seen directly from the gauge conditions on the metric.

\end{itemize}

\section*{Acknowledgments}
The authors thank Zhoujian Cao, Marc Geiller, and Xiaoning Wu for useful discussions, and Marc Geiller again for pointing out the typos and errors in the manuscript. P.M. would like to thank Glenn Barnich for long term collaborations and supports in relevant
research topics. This work is supported in part by the National Natural Science Foundation of China (NSFC) under Grants No. 11935009 and No. 11905156.

\appendix

\section{Finite twisting asymptotic symmetry transformations}

In this Appendix, we work out the finite version of the asymptotic symmetry. In the NU gauge, this was explicitly derived by Barnich and Troessaert \cite{Barnich:2016lyg}. Here, we will follow the same procedure with relaxed gauge and fall-off conditions, see also \cite{Mao:2019ahc} for other but relevant generalizations. In the NP formalism, the local Lorentz transformation is described in three classes of rotation of the tetrad system \cite{Chandrasekhar}. A combined form of the three rotations is given by
\begin{align}
&\tilde{l}=(1+\bA B)(1 + A \bB)e^{E_R}l + B \bB e^{-E_R} n + \bB(1+\bA B)e^{iE_I}m + B(1+A\bB)e^{-iE_I}\bm,\nn\\
&\tilde{n}=A\bA e^{E_R}l + e^{-E_R}n + \bA e^{iE_I}m + A e^{-iE_I}\bm,\\
&\tilde{m}=A(1+\bA B)e^{E_R}l + Be^{-E_R}n + (1+\bA B)e^{iE_I} m + ABe^{-iE_I}\bm.\nn
\end{align}
A combined change of coordinates is in the form
\be
u=u(u',r',z',\bz'),\;\;r=r(u',r',z',\bz'),\;\;x^A=x^A(u',r',z',\bz').
\ee
The gauge conditions of $l$ imply that
\be
\frac{\p x^\mu}{\p r'}=\tilde l^\mu.
\ee
Hence,
\begin{align}
&\frac{\p u}{\p r'}=B \bB e^{-E_R} W + \bB(1+\bA B)e^{iE_I} M + B(1+A\bB)e^{-iE_I} \xbar M,\nn\\
&\frac{\p z}{\p r'}=X^z B \bB e^{-E_R}+L^z \bB(1+\bA B)e^{iE_I} + \bL^z B(1+A\bB)e^{-iE_I},\\
&\frac{\p r}{\p r'}=(1+\bA B)(1 + A \bB)e^{E_R}+U B \bB e^{-E_R} + \omega \bB(1+\bA B)e^{iE_I} + \xbar\omega B(1+A \bB)e^{-iE_I},\nn
\end{align}
which generalizes Eq. (6.41) of \cite{Barnich:2016lyg}. The unprimed coordinates will be fixed up to four integration constants of $r'$, for which we will denote with a subscript ``0''. To implement the gauge condition $\kappa=\pi=\epsilon=0$, one needs to check the transformed spin coefficients,
\be
\kappa'=-\tilde{l}^\nu \tilde{l}^\mu \nabla_\nu \tilde{m}_\mu,\;\;\xbar\pi'=-\tilde{l}^\nu \tilde{m}^\mu\nabla_\nu \tilde{n}_\mu,\;\;\epsilon'=\tilde{l}^\nu \tilde{n}^\mu\nabla_\nu \tilde{l}_\mu,
\ee
which eventually lead to
\begin{align}
&\tilde{D}B= e^{E} \left[B \bB e^{-E_R} \tau + \bB(1+\bA B)e^{iE_I}\sigma + B(1+A\bB)e^{-iE_I}\rho\right],\nn\\
&\tilde{D}\bA=\bA^2e^{E} \left[B \bB e^{-E_R} \tau + \bB(1+\bA B)e^{iE_I}\sigma + B(1+A\bB)e^{-iE_I}\rho\right]\nn\\
&\hspace{2cm}- e^{-E} \left[B \bB e^{-E_R} \nu + \bB(1+\bA B)e^{iE_I}\mu + B(1+A\bB)e^{-iE_I}\lambda\right],\label{Lorentz}\\
&\tilde{D}E=-2\bA e^{E} \left[B \bB e^{-E_R} \tau + \bB(1+\bA B)e^{iE_I}\sigma + B(1+A\bB)e^{-iE_I}\rho\right]\nn\\
&\hspace{2cm}- 2 \left[B \bB e^{-E_R} \gamma + \bB(1+\bA B)e^{iE_I}\beta + B(1+A\bB)e^{-iE_I}\alpha\right].\nn
\end{align}
where $\tilde{D}=\frac{\p}{\p r'}$ because of the gauge conditions on $l$. The following fall-off conditions of the symmetry parameters are assumed for asymptotically flat case \cite{Barnich:2016lyg}
\be\begin{split}
&r=e^{E_{R0}}r'+O(1),\;\;\;\;u=O(1),\;\;\;\;x^A=O(1),\\
&A=O(1),\;\;\;\;E=O(1),\;\;\;\;B=O(r^{-1}).
\end{split}\ee
Hence, the three classes of tetrad rotation will be fixed up to six integration constants of $r'$ at their leading order.
To proceed, we use the fall-off conditions of the NP variables to fix the integration constants. We start from the transformed tetrad $n'$, 
\be
n'^{u'}=e^{-E_{R0}}\p_u u'_0 + \cO(r^{-1}),\quad n'^{r'}=\cO(r),\quad 
n'^{z'}=e^{-E_{R0}}\p_u z'_0 + \cO(r^{-1}),
\ee
which implies
\be
E_{R0}=\p_u u'_0,\quad  z'_0=Y(z,\bz).
\ee
The next one is $m'$,
\be
m'^{u'}=\cO(r^{-1}),\quad m'^{r'}= \cO(r^{-1}),\quad m'^{z'}=e^{i E_{I_0}}P\p_{\bz} z' + \cO(r^{-2}),\quad m'^{\bz'}=\cO(r^{-1}).
\ee
This yields $z_0'=Y(z)$. The condition $\Gamma_{213}=\cO(r^{-2})$ yields that
\be
A_0=(\gamma^0+\xbar\gamma^0) e^{-2E_{R_0}}B_0 + e^{-E_{R_0}}\eth_n' E_{R_0}.
\ee
The condition $Re(\rho)=\cO(r^{-3})$ fixes $r_0'$ as
\begin{multline}
r_0'= \frac12 \xbar A_0 B_0 + \frac12 A_0 \xbar B_0 + \frac12 e^{-E_{R_0}}(\xbar\eth_n' B_0 +  \eth_n' \xbar B_0)\\
-\frac12 e^{-E_{R_0}}\left(B_0 \xbar\eth_n' E_0 + \xbar B_0 \eth_n' \xbar E_0 + e^{E_{I_0}}\tau^0\xbar B_0 + e^{-E_{I_0}} \xbar \tau^0 B_0\right).
\end{multline}
There is no condition to fix $B_0$. This is the generalization of the Weyl-BMS transformation \cite{Barnich:2016lyg} by removing the hypersurface orthogonal condition.

\providecommand{\href}[2]{#2}\begingroup\raggedright\endgroup

\end{document}